\documentclass[conference]{IEEEtran}
\ifCLASSINFOpdf
  \usepackage[pdftex]{graphicx}
  \usepackage{dblfloatfix}
  \usepackage{fixltx2e}
  \graphicspath{{./figures/}}
  \DeclareGraphicsExtensions{.pdf,.jpeg,.png}
\else
\fi
\begin{document}
%
\title{High Performance and Scalable AWG for Superconducting Quantum Computer}

\author{\IEEEauthorblockN{Jin~Lin,
        Fu-Tian~Liang, 
        Yu~Xu,
        Li-Hua~Sun,
        Cheng~Guo,
        Sheng-Kai~Liao
        and Cheng-Zhi~Peng 
        }
\IEEEauthorblockA{Hefei National Laboratory for Physical Sciences
at the Microscale \\and Department of Modern Physics University of Science and Technology of China
Hefei, 230026, China\\
Chinese Academy of Sciences (CAS) Center for Excellence and Synergetic Innovation Center \\in Quantum Information and Quantum Physics, University of Science and Technology of China, Shanghai 201315, China\\
Email: ljt0132@ustc.edu.cn
}

\IEEEcompsocitemizethanks{\IEEEcompsocthanksitem Jin Lin, Fu-Tian Liang, Yu Xu, Li-Hua Sun, Cheng Guo, Sheng-Kai Liao and Cheng-Zhi Peng are with National Laboratory for Physical Sciences at the Microscale  
, University of Science and Technology of China, Hefei, Anhui 230026, P.R. of China.\protect\\
E-mail: see http://www.michaelshell.org/contact.html
 }
\thanks{Manuscript received April 19, 2018; revised August 26, 2018. This work is supported by the National Natural Science Foundation of China under grant number 61401422, 11375179, 11375263. }}


%


\maketitle

\begin{abstract}
Superconducting quantum computer is manufactured based on semiconductor process which makes qubits integration possible. At the same time, this kind of qubit exhibits high performance fidelity, de-coherence time, scalability and requires a programmable arbitrary waveform generator (AWG). This paper presents implementation of an AWG which composed of two gigabit samples per second (GSPS) sampling rate, 16 bit vertical resolution digital to analog converters (DACs). The AWG integrated with separate microwave devices onto a metal plate for the scale-up consideration. A special waveform sequence output controller is designed to realize seamless waveform switching and arbitrary waveform generator. The jitter in multiple AWG channels is around 10ps, Integral nonlinearity (INL) as well as differential nonlinearity (DNL) is about 2 LSB, and the qubit performance of the de-coherence time (T2*) achieved 33\% promotion over that of a commercial 1 GSPS, 14 bit AWG.
\end{abstract}


%
\IEEEpeerreviewmaketitle

\section{Introduction}
Superconducting qubit represents one of the most possible quantum computing schemes, which exhibits excellent performance in fidelity, de-coherence time and integration. Qubit controlling and reading can be achieved by commercial 1 GSPS DAC and ADC by use of IQ mixer in microwave modulation for up/down conversion \cite{chen2012multiplexed,ofek2016extending}.

\begin{figure}[!t]
\centering
\includegraphics[width=3.8in]{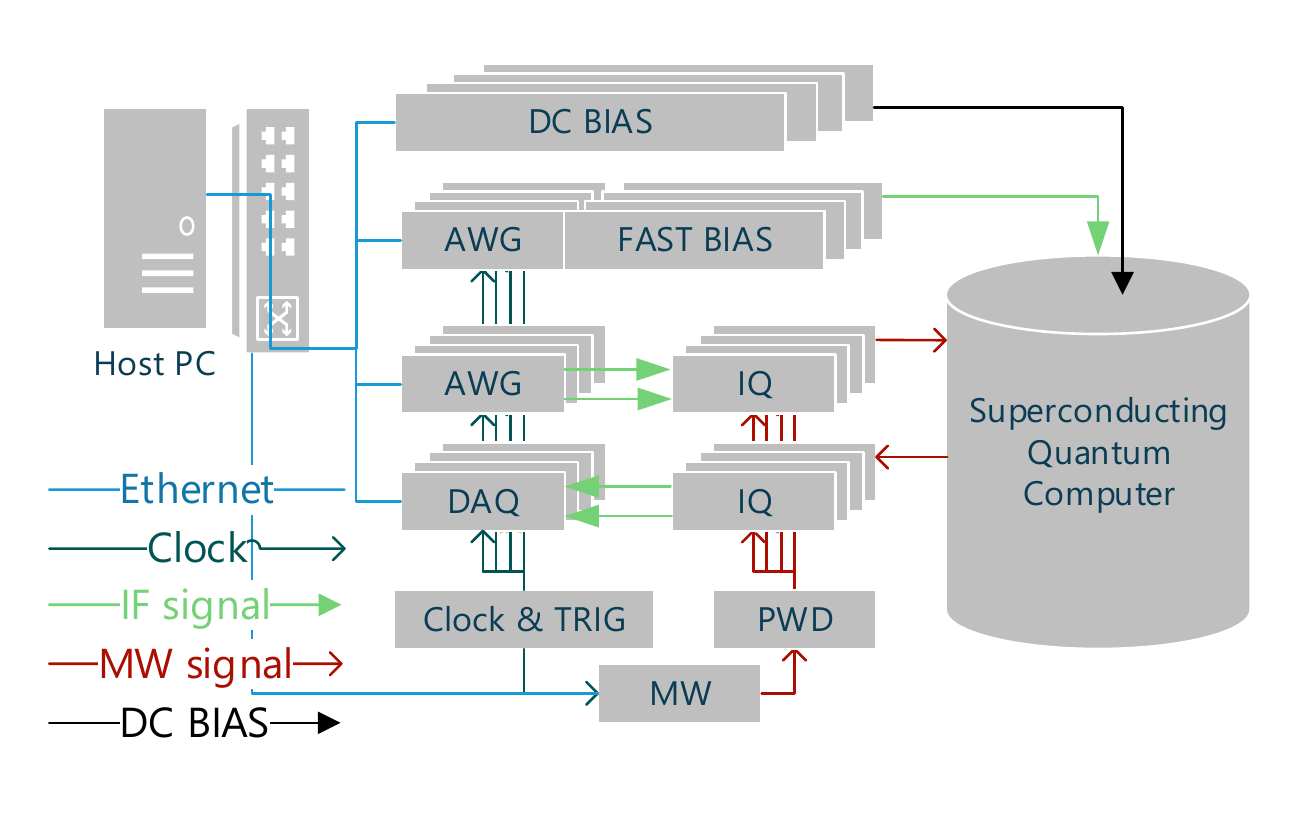}
\caption{Block diagram of superconducting quantum processor control system.}
\label{chap3_sqc_control_plan}%
\end{figure}

As shown in Fig.~\ref{chap3_sqc_control_plan}, superconducting quantum computing control system is composed of multiple independent parts, including switching network, clock \& synchronization, qubit control, qubit read, bias and host. Each control unit is formed by AWGs, filters, differential amplifiers, power splitter and high precision DC sources. These parts can be used flexibly in accordance with the needs of system control requirement. Readout modules include a data acquisition (DAQ) and other parts that included in control unit. The clock and trigger system consists of a phase locked loop (PLL) and multiple signal fan-out units that supports the function of fanning out one input to multiple outputs, providing the clock and trigger signals that required for system synchronization.

To perform a qubit measurement, the control module prepares the qubits to a known state, and then, the readout module reads out the information of the qubits in a nondestructive measurement mode. Next, according to the qubit's state information, the control module carries out the feedback control on AWGs' output.

In general, a quantum chip is manufactured with each qubit exhibits different characteristic frequencies. As a result, one may concurrently read multiple qubits by frequency superposition in one read out module. Typically, ratio between the channel number of qubit control and qubit readout is around 4 to 10. This means the proportion between the channel number of AWG and DAQ is about 10 to 1. Hence, high-speed, high-precision AWG and its synchronization control represent the demand and challenges in such a system. The required synchronization accuracy between the channels of superconducting qubits is in the order of picoseconds. 

The design of superconducting quantum bit chip is constantly exploring, and its new control requirements are springing up. For example, rapid calibration of modulation waveform, realization of specific rapid demodulation algorithm, and rapid feedback control required by the fault-tolerant quantum computing etc. Such a control device represents a highly customized system. 

\begin{figure*}[t]
  \centering
  \includegraphics[width=0.9\linewidth]{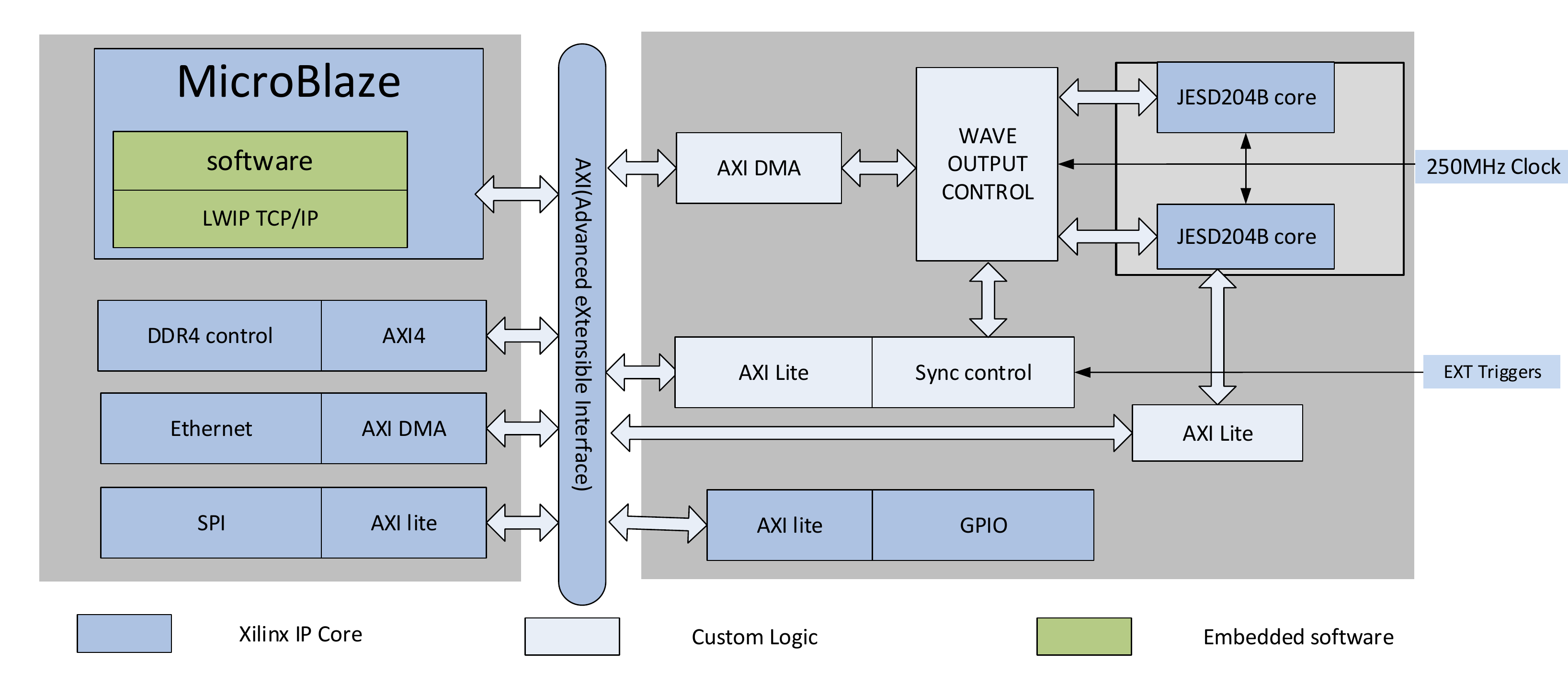}
  \caption{Block diagram of AWG FPGA logic.}\label{chap3_logic_frame}
\end{figure*}
Over the past few years, several groups, e.g. in UCSB, IBM, ETH Zurich, Yale and BBN technologies, have implemented superconducting quantum computing control platforms\cite{kaufmann2013dac, Castelvecchi2017IBM,ryan2017hardware}. Moreover, control systems have been realized in ultralow temperature environments\cite{homulle2016cryogenic, homulle2017reconfigurable}. However, commercially available devices cannot satisfy all the requirements, especially for programmable and compact demands. Therefore, to realize a large scale quantum computer, it is desirable to develop a high performance programmable, compact and extensible arbitrary waveform generator (AWG). 
This paper presents an AWG framework and waveform output sequence controller with extensible consideration. To validate the self-made AWG, we compare the qubit de-coherence time T2* measurement results of our AWG with those of the commercial 1 GSPS 14 bit AWG. 

\section{AWG IMPLEMENTATION}

\subsection{HARDWARE}
\begin{figure}[h]
\centering
\includegraphics[width=.5\textwidth]{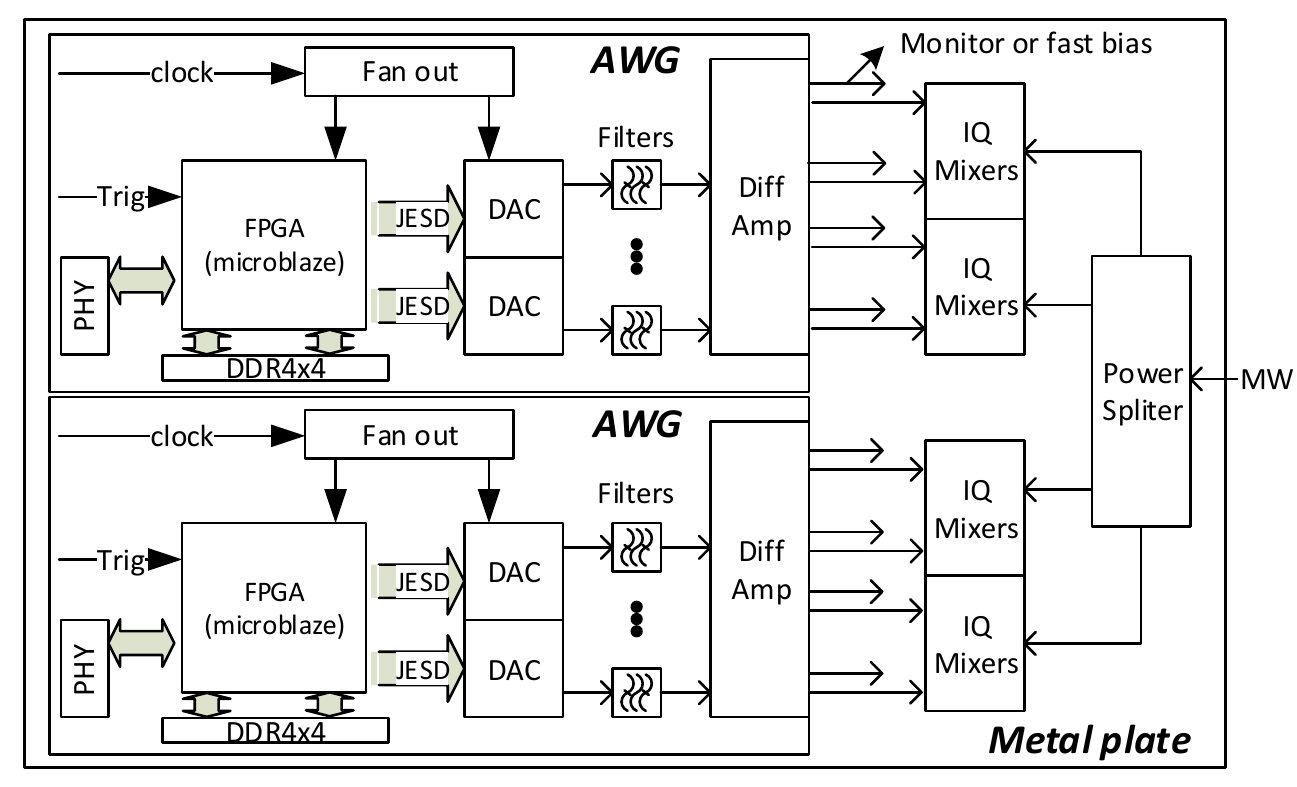}
\caption{ AWG framework.}
\label{AWG_framework}
\end{figure}
The AWG unit is composed by a Xilinx Field-Programmable Gate Array (FPGA) and two high performance Digital-to-Analog Converter (DAC) chips. Such an AWG is capable of providing four arbitrary waveform output channels, each of which operating at 2 GSPS and provide outputs with 16 bit resolution. We use the JESD204B protocol to achieve transmission of the high speed digital data rate of about 160 Gbps. Bipolar signals are achieved by passing the outputs through low pass filters and differential amplifiers. As shown in Fig.~\ref{AWG_framework}. a reliable communication is ensured by the information exchange via a gigabit Ethernet running on TCP/IP protocol. An external input clock that enables multi-boards to operate under a synchronous source is fanned out and distributed equally to the FPGA and DACs. Another external input trigger which is synchronized to the same clock source can provide accurate output control. An extensible AWG array may be easily achieved by integration of multiple AWGs with a dedicated synchronize control module. 


To reduce the size of the system, we integrate the AWG with individual IQ mixers and power splitters onto one metal plate, and constraint the height to 1U of the standard server rack. As shown in Fig. \ref{AWG_framework}, two channels are used from the same chip as I/Q signals and are wired to the IQ mixer’s corresponding inputs. A common microwave source is connected to the power splitter and four outputs are separately fanned out to 4 IQ mixers’ LO input. Every compensative channel exiting from the differential amplifier will be used for monitoring or fast bias controlling.

\subsection{FPGA IMPLEMENTATION}

\begin{figure*}[t]
  \centering
  \includegraphics[width=0.95\linewidth]{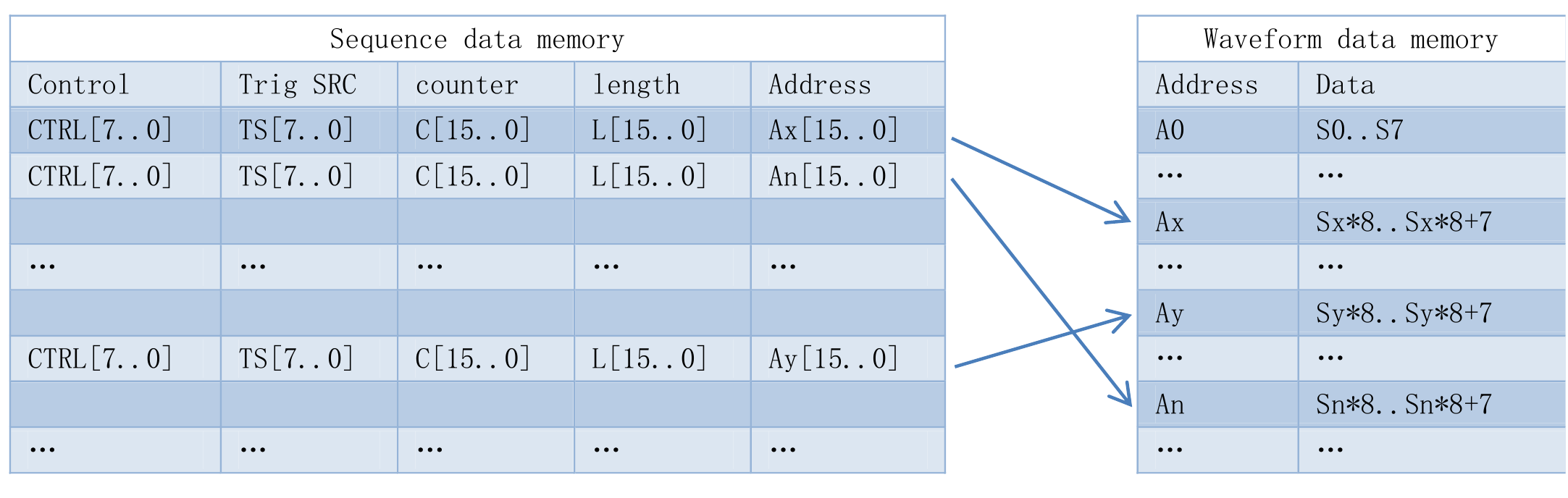}
  \caption{Sequence data and waveform data format.}\label{seq_format}
\end{figure*}

Figure.~\ref{chap3_logic_frame} shows the block diagram of the signal flow in the FPGA. 
A Xilinx MicroBlaze processor generated by the Vivado software is embedded in the FPGA and communicates with all peripherals through the Advanced eXtensible Interface (AXI) bus. 
The software running on the processor implements a reliable communication with the host computer through lightweight TCP/IP stack. It also configures the operating parameters through the SPI/GPIO IPs on the AXI bus to the peripheral chips and function modules. Meanwhile, the working state of the AWG is periodically sent to local area network (LAN) through UDP broadcast packets, so that the system states may be monitored by any host in the same LAN. The key waveform output control data is also transmitted through the network and sent to the waveform output control module. 

The waveform output control module operates under the 250MHz clock which coming from the external clock interface. The same clock input makes multiple AWG modules operating in one clock source. The module interface with the DAC chip is completed by the JESD204B IP core which is operating in 10Gbps rate. The JESD204B protocol supports deterministic delay control, and realizes computable time delay of the waveform data from the JESD204B IP core input to the DAC output. 

\subsection{AWG CONTROL}
We implement a waveform output sequence controller for each channel. The controller is composed of three parts, i.e., a waveform data memory (WDM), a sequence data memory (SDM) and a finite state machine (FSM) for reading and controling. As shown in Fig.~\ref{seq_format}, the memory data may be set by the host-computer and provides direct arbitrary waveform output. According to the flag bits of the sequence data, the FSM reads the sequence data from the WDM and determines the output mode of the waveform data. Output control mode includes the start address, length, trigger source, counter value, etc. In general, the programmable arbitrary waveform generate function is achieved by the FSM together with the WDM and SDM. 

When running an arbitrary waveform output, at the beginning, the resources are ready for the first sequence data to execute. When a sequence data is running, the FSM prefetching the next sequence data to determine and prepare the resources required for its running. Hence, the resources are ready for every sequence data to be executed, and we may seamlessly switch to the next instruction without delay and splicing multiple waveform data. However, there is a constraint, where the output length of each waveform area cannot be less than 4. The purpose of doing this is to ensure that the prefetching action is accomplished in sufficient time to prepare the resources. 


\section{TESTING }

\subsection{INL DNL}
The AWG possesses 16 bit vertical resolution, that is, 65536 codes. The INL/DNL test requires traversing all the codes. We set a digital code and use a high-precision multimeter to measure the output voltage. Then, we continuously increase the digital code until all the codes are traversed. The test takes about 0.1 seconds for one code and about two hours to go through all the codes of a single channel. The high precision multimeter supports an external trigger and possesses 50k sample storage. We considered AWG with 65535 steps with the duration of 1ms for each step, which is synchronized to a trigger signal. The trigger signal is connected to the external trigger input of the multimeter. To collect the measured data, the host computer controls the AWG and the multimeter, and carries out the INL and DNL analysis. By improving the testing scheme, the test time of a single code value is reduced to 1ms, and the whole test time is reduced to 1 minute. 
The results of the test showing that the AWG DNL and INL are within 2LSB. 

\subsection{Phase noise}

\begin{figure}[h]
\centering
\includegraphics[width=.5\textwidth]{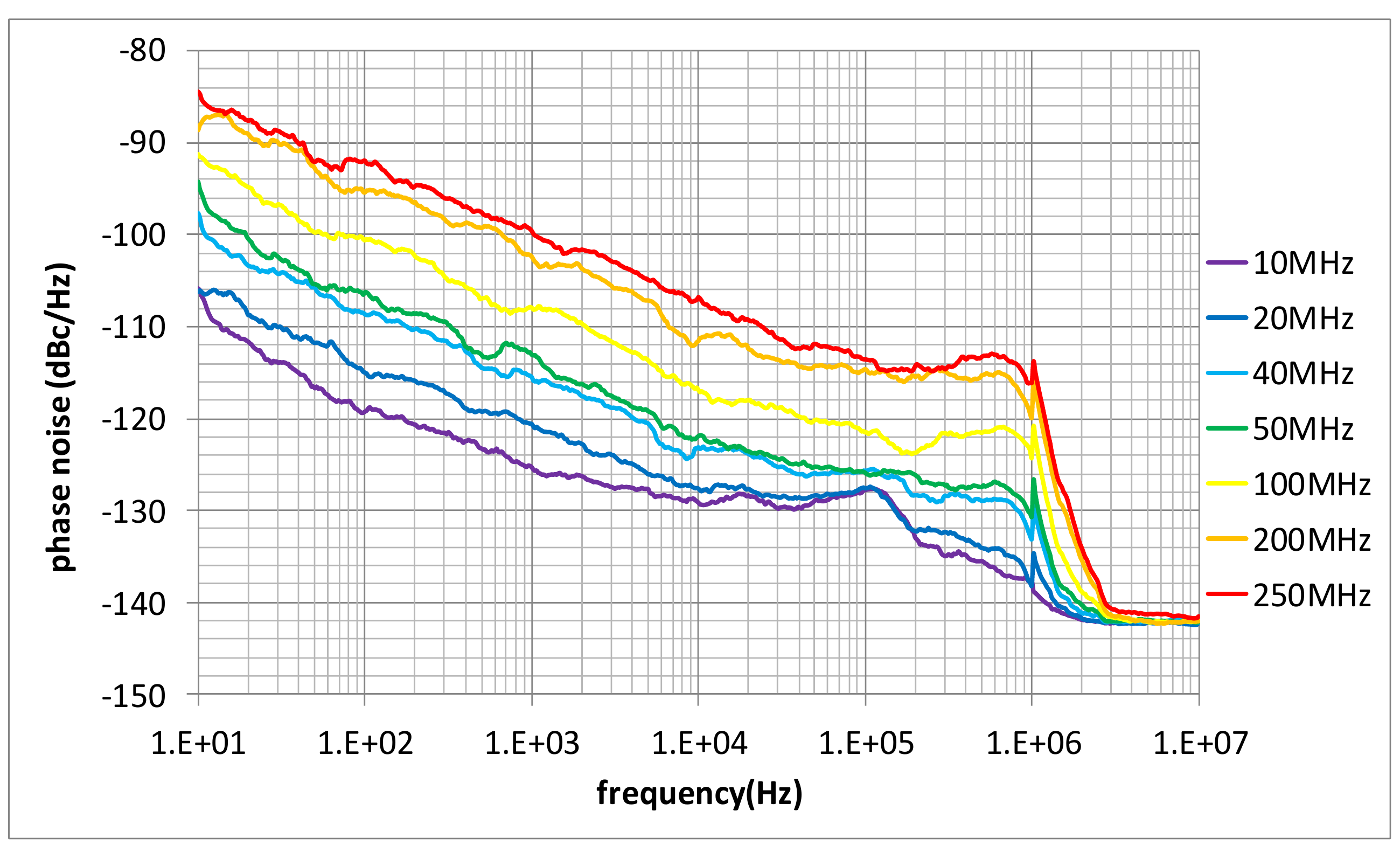}
\caption{Phase noises at different frequency of the AWG output}
\label{chap4_dac_dif_feq_phase_noise}
\end{figure}
Fig.\ref{chap4_dac_dif_feq_phase_noise} shows the output phase noise of one AWG channel at seven distinct frequencies, i.e., 10, 20, 40, 50, 100, 200 and 250MHz. The noise floor larger than 10MHz frequency offset tends to be consistent. Increasing the signal frequency yields movement of the phase noise curve. In particular, the phase noise curve moves \~6dB when the frequency is doubled, which is consistent with the theory.

\subsection{Spurious free dynamic range (SFDR)}
\begin{figure}[h]
\centering
\includegraphics[width=.5\textwidth]{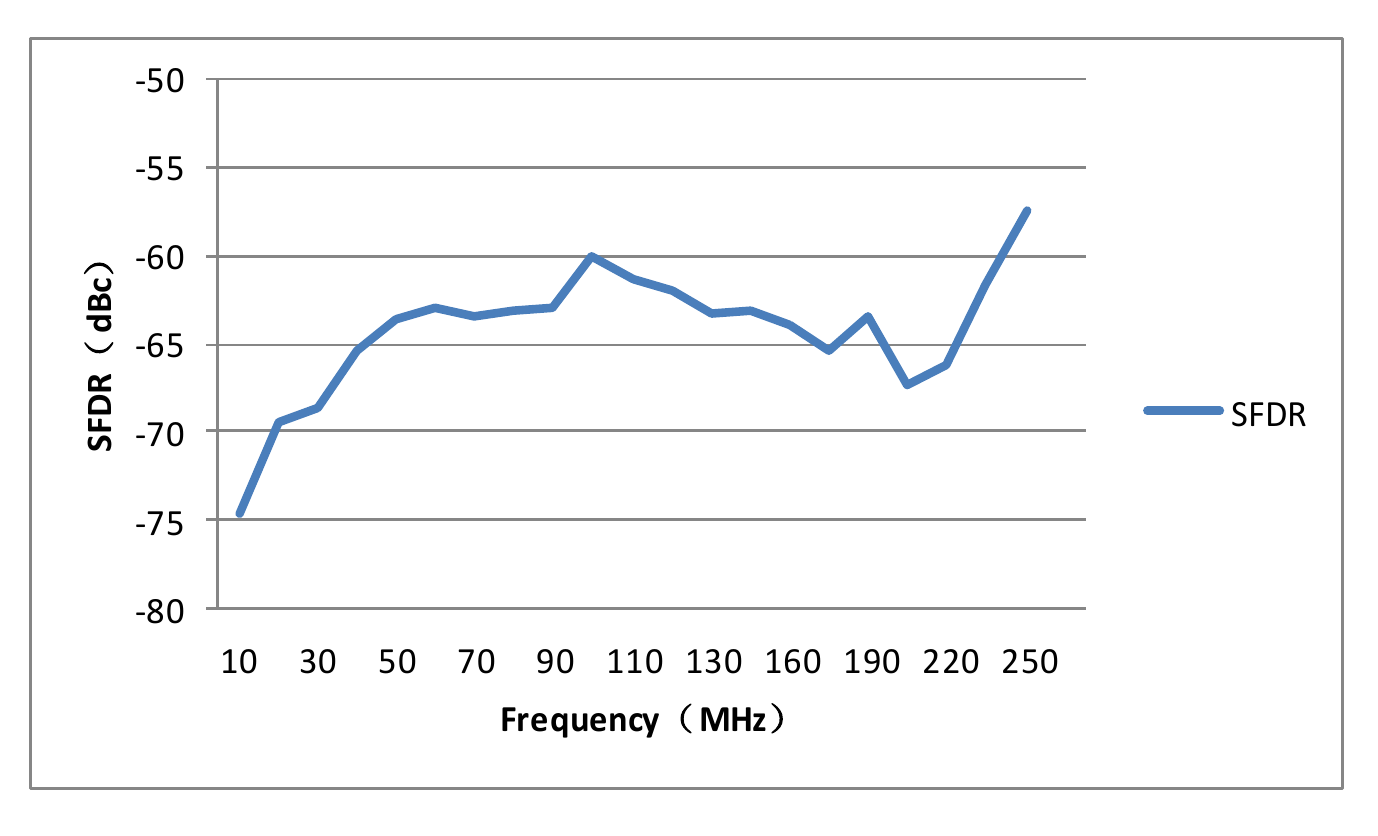}
\caption{SFDR test result}
\label{chap4_sfdr}
\end{figure}
In the SFDR measurement, the harmonic noise represents the main spurious noise source. Consider measurement of the harmonic noise, especially when the measurement signal is a monosyllabic sine wave signal. If the RF attenuation setting of the spectrum meter is not reasonable, considerable difference between the tested harmonic noise and the actual harmonic noise will be observed. We consider the lowest output as the gain of the AWG, and set the frequency spectrum instrument input attenuation to 0dB. As shown in Fig.\ref{chap4_sfdr}, the SFDR curve is achieved at 25 frequency points, i.e., from 10MHz to 250MHz with 10MHz step size. The results are consistent with the chip’s datasheet.


\subsection{SYNCHRONIZATION}

\begin{figure}[h]
\centering
\includegraphics[width=.5\textwidth]{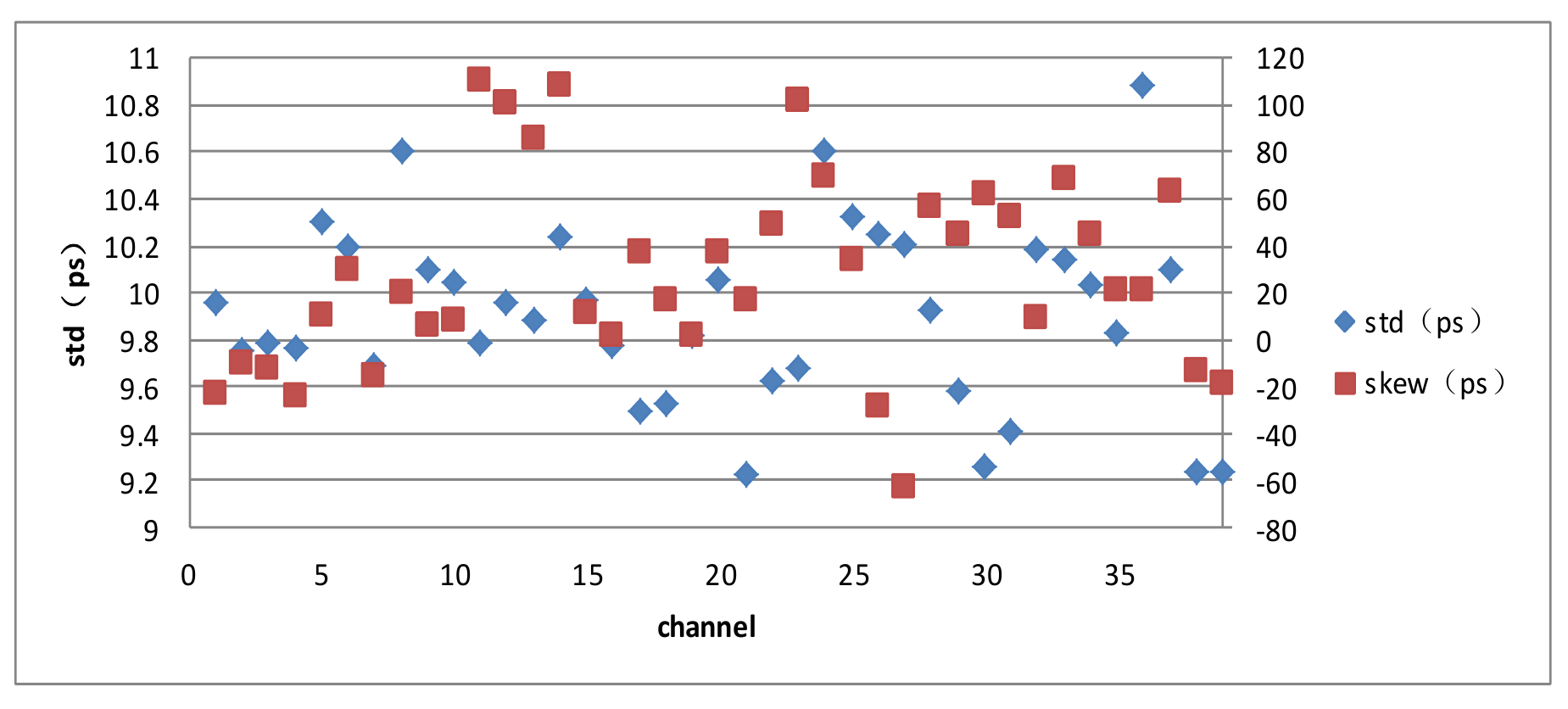}
\caption{Jitters among multipule AWG channels}
\label{chap4_wave_sync}
\end{figure}
Fig.\ref{chap4_wave_sync} shows the jitters of 40 arbitrary waveform output channels on 10 AWGs. The minimum and maximum standard deviations are 9.22ps and 10.89ps, respectively, with the mean value of 9.9ps. The skew between the channels is ˜100ps. The large skew between the channels is due to the fact that different signal line delay amount the clock signals input to the AWGs, and the different signal line delay from the AWG output to the oscilloscope input. The skew is deterministic and may be adjusted by cable length. 

\subsection{QUBIT TEST}
\begin{figure}[h]
\resizebox{9cm}{!}{\includegraphics[scale=1]{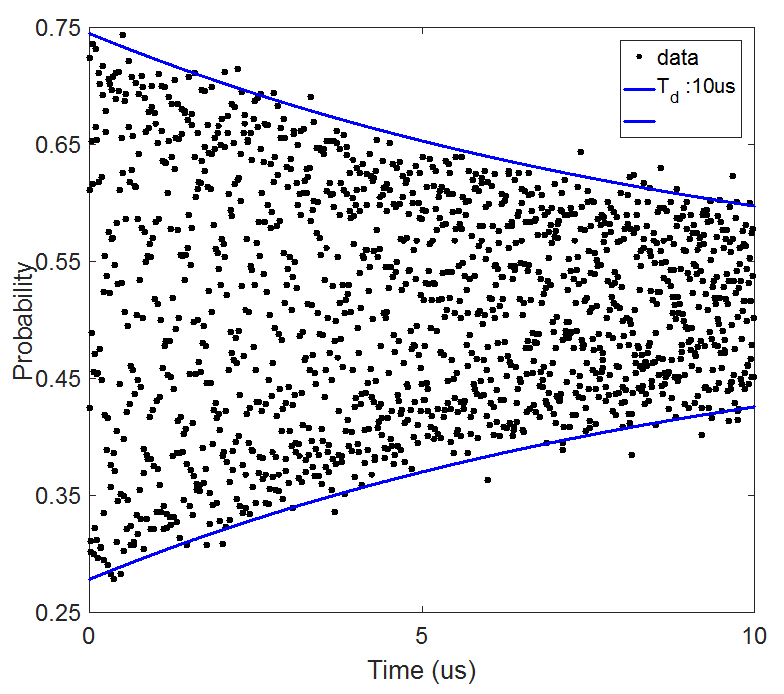}}
\caption{T2* measurement results  for commercial AWG.\label{T2_awg}}%
\end{figure}

\begin{figure}[h]
\resizebox{9cm}{!}{\includegraphics[scale=1]{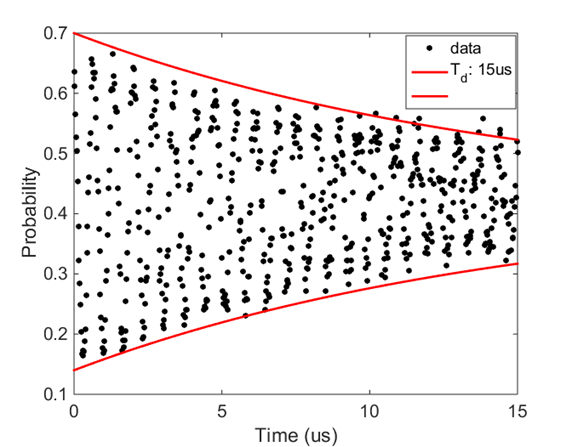}}
\caption{T2* measurement results  for self-made AWG\label{T2_ustc}}%
\end{figure}
The implemented AWG is tested by a $7_{1/2}$ voltmeter and a 1 GHz bandwidth oscillator. It is demonstrated that the AWG exhibits 14 effective bits, and 10ps jitter among different channels. We test the AWG’s performance further by measuring the qubit's de-coherence time (T2*) which representing an important parameter for qubits~\cite{bialczak2011development,song201710}. Fig.\ref{T2_awg} and Fig.\ref{T2_ustc} compare the results driven by commercial 1 GSPS, 14 bit AWG. As shown in the figure, T2* of 10us and 15us is achieved separately driven by commercial AWG and our AWG, which is about 33\% of the performance promotion. 

\section{Conclusion}
We introduced a scalable and highly integrated AWG array for the superconducting quantum computing control system. The AWG consists of two DACs, one FPGA and a gigabit Ethernet transceiver. An array has been used for the superconducting quantum computing control system. We also investigated the performance of the proposed AWG in detail and compared the qubit T2* test results to that of commercial AWG. 


%
%



\bibliographystyle{IEEEtran}
\bibliography{IEEEabrv,./bib/paper}

\begin{thebibliography}{1}
\providecommand{\url}[1]{#1}
\csname url@samestyle\endcsname
\providecommand{\newblock}{\relax}
\providecommand{\bibinfo}[2]{#2}
\providecommand{\BIBentrySTDinterwordspacing}{\spaceskip=0pt\relax}
\providecommand{\BIBentryALTinterwordstretchfactor}{4}
\providecommand{\BIBentryALTinterwordspacing}{\spaceskip=\fontdimen2\font plus
\BIBentryALTinterwordstretchfactor\fontdimen3\font minus
  \fontdimen4\font\relax}
\providecommand{\BIBforeignlanguage}[2]{{%
\expandafter\ifx\csname l@#1\endcsname\relax
\typeout{** WARNING: IEEEtran.bst: No hyphenation pattern has been}%
\typeout{** loaded for the language `#1'. Using the pattern for}%
\typeout{** the default language instead.}%
\else
\language=\csname l@#1\endcsname
\fi
#2}}
\providecommand{\BIBdecl}{\relax}
\BIBdecl

\bibitem{chen2012multiplexed}
Y.~Chen, D.~Sank, P.~O'Malley, T.~White, R.~Barends, B.~Chiaro, J.~Kelly,
  E.~Lucero, M.~Mariantoni, A.~Megrant \emph{et~al.}, ``Multiplexed dispersive
  readout of superconducting phase qubits,'' \emph{Applied Physics Letters},
  vol. 101, no.~18, p. 182601, 2012.

\bibitem{ofek2016extending}
N.~Ofek, A.~Petrenko, R.~Heeres, P.~Reinhold, Z.~Leghtas, B.~Vlastakis, Y.~Liu,
  L.~Frunzio, S.~Girvin, L.~Jiang \emph{et~al.}, ``Extending the lifetime of a
  quantum bit with error correction in superconducting circuits,''
  \emph{Nature}, vol. 536, no. 7617, p. 441, 2016.

\bibitem{kaufmann2013dac}
T.~Kaufmann, T.~J. Keller, J.~M. Franck, R.~P. Barnes, S.~J. Glaser, J.~M.
  Martinis, and S.~Han, ``Dac-board based x-band epr spectrometer with
  arbitrary waveform control,'' \emph{Journal of Magnetic Resonance}, vol. 235,
  pp. 95--108, 2013.

\bibitem{Castelvecchi2017IBM}
D.~Castelvecchi, ``Ibm's quantum cloud computer goes commercial.''
  \emph{Nature}, vol. 543, no. 7644, p. 159, 2017.

\bibitem{ryan2017hardware}
C.~A. Ryan, B.~R. Johnson, D.~Rist{\`e}, B.~Donovan, and T.~A. Ohki, ``Hardware
  for dynamic quantum computing,'' \emph{Review of Scientific Instruments},
  vol.~88, no.~10, p. 104703, 2017.

\bibitem{homulle2016cryogenic}
H.~Homulle, S.~Visser, and E.~Charbon, ``A cryogenic 1 gsa/s, soft-core fpga
  adc for quantum computing applications,'' \emph{IEEE Transactions on Circuits
  and Systems I: Regular Papers}, vol.~63, no.~11, pp. 1854--1865, 2016.

\bibitem{homulle2017reconfigurable}
H.~Homulle, S.~Visser, B.~Patra, G.~Ferrari, E.~Prati, F.~Sebastiano, and
  E.~Charbon, ``A reconfigurable cryogenic platform for the classical control
  of quantum processors,'' \emph{Review of Scientific Instruments}, vol.~88,
  no.~4, p. 045103, 2017.

\bibitem{bialczak2011development}
R.~R.~C. Bialczak, \emph{Development of The Fundamental Components of A
  Superconducting Qubit Quantum Computer}.\hskip 1em plus 0.5em minus
  0.4em\relax University of California, Santa Barbara, 2011.

\bibitem{song201710}
C.~Song, K.~Xu, W.~Liu, C.-p. Yang, S.-B. Zheng, H.~Deng, Q.~Xie, K.~Huang,
  Q.~Guo, L.~Zhang \emph{et~al.}, ``10-qubit entanglement and parallel logic
  operations with a superconducting circuit,'' \emph{Physical review letters},
  vol. 119, no.~18, p. 180511, 2017.

\end{thebibliography}
%
%
%

\end{document}